\documentclass[aps,prl,twocolumn,showpacs,byrevtex]{revtex4}

\usepackage{times,mathptm}
\usepackage[dvips]{graphicx,color}

\begin{document}

\title{Ferrimagnetic Slater Insulator Phase of the Sn/Ge(111) Surface}
\author{Jun-Ho Lee, Hyun-Jung Kim, and Jun-Hyung Cho$^{*}$}
\affiliation{Department of Physics and Research Institute for Natural Sciences, Hanyang University,
17 Haengdang-Dong, Seongdong-Ku, Seoul 133-791, Korea}

\date{\today}

\begin{abstract}
We have performed the semilocal and hybrid density-functional theory (DFT) studies of the Sn/Ge(111) surface to identify the origin of the observed insulating ${\sqrt{3}}{\times}{\sqrt{3}}$ phase below ${\sim}$30 K. Contrasting with the semilocal DFT calculation predicting a metallic 3${\times}$3 ground state, the hybrid DFT calculation including van der Waals interactions shows that the insulating ferrimagnetic structure with ${\sqrt{3}}{\times}{\sqrt{3}}$ structural symmetry is energetically favored over the metallic 3${\times}$3 structure. It is revealed that the correction of self-interaction error with a hybrid exchange-correlation functional gives rise to a band-gap opening induced by a ferrimagnetic order. The results manifest that the observed insulating phase is attributed to the Slater mechanism via itinerant magnetic order rather than the hitherto accepted Mott-Hubbard mechanism via electron correlations.
\end{abstract}


\maketitle


Two-dimensional (2D) electronic systems formed at crystal surfaces have attracted much attention because of their intriguing physical phenomena such as charge density waves (CDW), magnetic order, Mott insulators, and 2D superconductivity~\cite{tos,wei,car1}. One of the most popular quasi-2D systems is the ${\sqrt{3}}{\times}{\sqrt{3}}$ phase formed by the 1/3-monolayer adsorption of group IV metal atoms, Sn or Pb, on the Si(111) or Ge(111) surface~\cite{pro,tej2,col,mor,col2,li,han,mod,ron2}. Here, adatoms locating at $T_4$ sites [Fig. 1(a)] saturate all the dangling bonds (DBs) of Si or Ge surface atoms, leaving a single DB for each adatom~\cite{nor}. Since these DB electrons are separated as far as ${\sim}$7 {\AA}, the resulting narrow half-filled DB band is likely to invoke various instabilities and strong electron correlations.

We here focus on a prototypical example of quasi-2D systems, Sn/Ge(111). Below ${\sim}$220 K, this system undergoes a reversible phase transition from a ${\sqrt{3}}{\times}{\sqrt{3}}R$30$^{\circ}$  (hereafter designated as ${\sqrt{3}}{\times}{\sqrt{3}}$) structure to a 3${\times}$3 structure~\cite{car2}. This phase transition was initially interpreted in terms of a surface CDW formation stabilized by electron correlation effects in the ${\sqrt{3}}{\times}{\sqrt{3}}$ structure~\cite{car2,gol}. However, the high-temperature ${\sqrt{3}}{\times}{\sqrt{3}}$ phase was later explained as a dynamical effect in which inequivalent Sn atoms interchange their vertical positions~\cite{uhr,avi,ron}: i.e., the three Sn atoms [up (U) and down (D) atoms in Fig. 1(b)] of two different heights within the 3${\times}$3 structure fluctuate between two positions as temperature increases, apparently showing a ${\sqrt{3}}{\times}{\sqrt{3}}$ structural symmetry~\cite{ron}. Nevertheless, the ground state of Sn/Ge(111) is still subject to much debate because, as the temperature further decreases to ${\sim}$30 K, various experimental techniques such as scanning tunneling microscopy (STM), low-energy electron diffraction, and angle-resolved photoemission spectroscopy (ARPES) explored a phase transition from the 3${\times}$3 structure to a new ${\sqrt{3}}{\times}{\sqrt{3}}$ structure where three Sn atoms in 3${\times}$3 unit cell have an equivalent height~\cite{cor}. This structural phase transition was observed to accompany simultaneously a metal-insulator transition (MIT)~\cite{cor}. However, previous density-functional theory (DFT) studies~\cite{gir,gor,flo,pul,ort,soft} with the local density approximation (LDA) and generalized gradient approximation (GGA) predicted that the ${\sqrt{3}}{\times}{\sqrt{3}}$ structure was not only metallic but also less stable than the 3${\times}$3 structure, thereby failing to describe the electronic and energetic properties of the observed insulating ${\sqrt{3}}{\times}{\sqrt{3}}$ phase. To resolve this problem of LDA, Profeta and Tosatti~\cite{pro} took into account Coulomb interactions (Hubbard $U$) using the LDA + $U$ scheme. They found that electron correlations can stabilize a magnetic insulator with ${\sqrt{3}}{\times}{\sqrt{3}}$ structural symmetry, indicating a Mott-Hubbard insulating ground state. On the other hand, Flores $et$ $al.$ pointed out that the effect of electron correlations cannot induce a transition to a Mott insulating ground state~\cite{flo}. Thus, it remains elusive whether the formation of the insulating phase of Sn/Ge(111) is attributed to electron correlations~\cite{tej1}.

\begin{figure}[ht]
\centering{ \includegraphics[width=7.0cm]{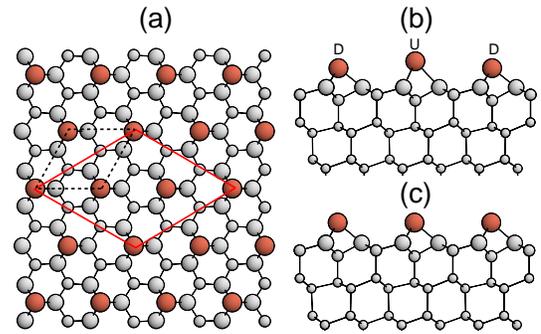} }
\caption{(Color on line) (a) Top and (b) side views of the metallic 3${\times}$3 structure of Sn/Ge(111). The side view of the insulating ferrimagnetic structure is given in (c). The dark and gray circles represent Sn and Ge atoms, respectively. U and D in (b) represent the up and down Sn atoms, respectively. The Sn atoms are located at the $T_4$ site, i.e., a single threefold hollow site above a second layer Ge atom. For distinction, Ge atoms in the subsurface layers are drawn with small circles. In (a), the 3${\times}$3 and ${\sqrt{3}}{\times}{\sqrt{3}}$ unit cells are indicated by the solid and dashed lines, respectively. The {\bf x} and {\bf z} directions in (b) are [11${\overline{2}}$] and [111], respectively.}
\end{figure}

In the present work, we propose a new origin of the observed insulating phase in Sn/Ge(111) due to itinerant magnetism. This magnetically driven insulating phase via an itinerant single-electron approach is characterized as a Slater insulator~\cite{slater}. It is found that the DB states of Sn atoms exhibit an itinerant character because of their significant hybridization with the Ge surface states, differing from the case of the previously proposed Mott-Hubbard insulator~\cite{pro} where each DB electron was treated to be localized at Sn adatom site. We note that the LDA and GGA tend to stabilize artificially delocalized electronic states due to their inherent self-interaction error (SIE), because delocalization reduces the spurious self-repulsion of electron~\cite{sie1,sie2}. In this regard, previous LDA and GGA calculations for Sn/Ge(111) may overestimate the stability of the metallic 3${\times}$3 structure~\cite{gir,pul,ort,flo,gor,soft}. It is, therefore, very interesting to examine if the observed insulating phase can be predicted by the correction of SIE with an exchange-correlation functional beyond the LDA or GGA.

In this Letter, we present a new theoretical study for Sn/Ge(111) based on the hybrid DFT scheme including van der Waals (vdW)~\cite{vdw} interactions (termed DFT+vdW scheme). We find that the correction of SIE with the hybrid exchange-correlation functional of Heyd-Scuseria-Ernzerhof (HSE)~\cite{hse} stabilizes the insulating ferrimagnetic (FI) structure where the band-gap opening occurs by a FI spin ordering of three Sn atoms within the 3${\times}$3 unit cell. Here, the buckling of three Sn atoms is suppressed to show apparently a ${\sqrt{3}}{\times}{\sqrt{3}}$ structural symmetry, and the stability of the FI structure is further enhanced by the inclusion of vdW interactions. The calculated magnetic moment of the FI structure is well distributed over Sn adatoms as well as Ge substrate atoms, giving rise to a magnitude of 1 ${\mu}_{\rm B}$ per 3${\times}$3 unit cell. It is thus demonstrated that the observed insulating phase in Sn/Ge(111) can be represented as a Slater insulator through itinerant magnetism, not as the previously~\cite{pro} proposed Mott-Hubbard insulator by Coulomb interactions $U$.

The present semilocal and hybrid DFT calculations were performed using the FHI-aims~\cite{aims} code for an accurate, all-electron description based on numeric atom-centered orbitals, with ``tight" computational settings. For the exchange-correlation energy, we employed the GGA functional of Perdew-Burke-Ernzerhof (PBE)~\cite{per} as well as the hybrid functional of HSE~\cite{hse}. The ${\bf k}$-space integration was done with the 15${\times}$15 and 9${\times}$9 uniform meshes in the surface Brillouin zones of the ${\sqrt{3}}{\times}{\sqrt{3}}$ and 3${\times}$3 unit cells, respectively. The Ge(111) substrate was modeled by a 6-layer slab with ${\sim}$34 {\AA} of vacuum in between the slabs~\cite{ctest}. Here, we used the optimized Ge lattice constants $a_0$ = 5.783, 5.718, and 5.667 {\AA} for the PBE, HSE, and HSE+vdW calculations, respectively. The HSE+vdW lattice constant~\cite{mar,zha} agrees most with the experimental value of 5.658 {\AA}~\cite{expt1}. Each Ge atom in the bottom layer was passivated by one H atom. All atoms except the bottom layer were allowed to relax along the calculated forces until all the residual force components were less than 0.02 eV/{\AA}. The employed HSE+vdW scheme was successfully applied to determine the energy stability of the metallic and insulating phases in indium nanowires on Si(111)~\cite{kim}.

\begin{table}[ht]
\caption{Calculated total energies (in meV per ${\sqrt{3}}{\times}{\sqrt{3}}$ unit cell) of the 1U2D, FM, and FI structures relative to the NM ${\sqrt{3}}{\times}{\sqrt{3}}$ structure. For comparison, the previous LDA and GGA results are also given.}
\begin{ruledtabular}
\begin{tabular}{lccc}
            &      1U2D    &    FM      &       FI  \\ \hline
PBE         &      $-$10.1     &        $-$  &      $-$   \\
HSE+vdW     &      $-$31.6     &    $-$18.7     &   $-$37.7  \\
LDA (Ref. ~\cite{ort}) &      $-$5     &        $-$  &      $-$   \\
LDA (Ref. ~\cite{pul}) &      $-$7.5     &        $-$  &      $-$   \\
LDA (Ref. ~\cite{pro}) &      $-$9     &        $-$  &      $-$   \\
GGA (Ref. ~\cite{avi}) &      $-$5     &        $-$  &      $-$   \\
\end{tabular}
\end{ruledtabular}
\end{table}

\begin{figure}[ht]
\centering{ \includegraphics[width=7.0cm]{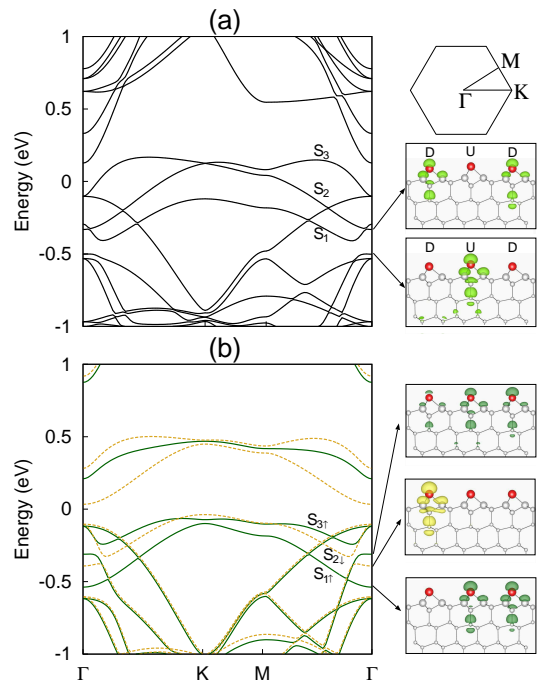} }
\caption{(Color on line) (a) Surface band structure of the 1U2D structure computed using the PBE functional. The spin-polarized surface band structure of the FI structure computed using the HSE+vdW scheme is given in (b). The band dispersions are plotted along the symmetry lines of the Brillouin zone of the 3${\times}$3 unit cell [see the inset in (a)]. The ${\Gamma}$-M line corresponds to [11${\overline{2}}$] direction. The charge characters of the DB states at the ${\Gamma}$ point are also displayed with an isosurface of 0.006 e/{\AA}$^3$. The energy zero represents the Fermi level. The majority and minority bands in (b) are drawn with the solid and dashed lines, respectively.}
\end{figure}

We begin to optimize the nonmagnetic (NM) ${\sqrt{3}}{\times}{\sqrt{3}}$ and 3${\times}$3 structures using the PBE functional. The optimized top and side views of the 3${\times}$3 structure are displayed in Fig. 1(a) and 1(b), respectively. It is seen that the U atom positions higher than the two D atoms by 0.35 {\AA}, in good agreement with those (ranging from 0.26 to 0.36 {\AA}) of previous DFT studies~\cite{soft,osk,ort,flo,gor,pul,pro,gir}. This so-called 1U2D structure is more stable than the NM ${\sqrt{3}}{\times}{\sqrt{3}}$ structure by 10.1 meV per ${\sqrt{3}}{\times}{\sqrt{3}}$ unit cell, which is well comparable with previous DFT results (see Table I). As shown in Fig. 1S of the Supplemental Material~\cite{supp} [Fig. 2(a)], the calculated band structure for the NM ${\sqrt{3}}{\times}{\sqrt{3}}$ (1U2D) structure exhibits the presence of occupied DB state(s) at the Fermi level, indicating a metallic feature. It is revealed that the charge characters of the DB states in the 1U2D structure [see Fig. 2(a)] represent a charge transfer from the D to the U atoms~\cite{pul}. This charge transfer gives rise to a reduced Coulomb repulsion between Sn DB electrons, resulting in a more stabilization over the NM ${\sqrt{3}}{\times}{\sqrt{3}}$ structure. Our spin-polarized PBE calculations were not able to find any spin ordering within the ${\sqrt{3}}{\times}{\sqrt{3}}$ and 3${\times}$3 unit cells. Thus, we can say that the semilocal DFT scheme with the PBE functional cannot predict the insulating ${\sqrt{3}}{\times}{\sqrt{3}}$ phase observed below ${\sim}$30 K~\cite{cor}.

It is noteworthy that a recent high-resolution photoemission study~\cite{tej2} for the Sn 4$d$ core level of the metallic 3${\times}$3 structure resolved three components, which were assigned to each of the three Sn atoms within the 3${\times}$3 unit cell. On the basis of this photoemission data together with STM images, Tejeda $et$ $al.$~\cite{tej2} concluded that the two D atoms position at slightly different heights, forming an inequivalent-down-atoms (IDA) structure. Our PBE calculation shows that the IDA structure with a height difference of ${\sim}$0.03 {\AA} between the two D atoms is almost degenerate in energy (less than 0.1 meV per Sn atom) with the 1U2D structure, consistent with a previous LDA calculation~\cite{osk}. For the 1U2D and IDA structures, we calculate the Sn 4$d$ core-level shifts using initial-state theory, where the shift is defined by the difference of the eigenvalues of the Sn 4$d$ core level at different sites. The results are displayed in Fig. 3 and compared to the photoemission experiment~\cite{tej2}. We find that each Sn 4$d$ core level is split into three sublevels, i.e., the degenerate $C_{d_1}$ ($C_{d_2}$) sublevel arising from the $d_{xy}$ and $d_{x^2-y^2}$ ($d_{yz}$ and $d_{xz}$) orbitals and the $C_{d_3}$ sublevel from the $d_{z^2}$ orbital. Note that such a crystal field splitting is conspicuous for the two D atoms but negligible for the U atom. On the basis of the calculated initial-state core levels, the observed $C_1$, $C_2$, and $C_3$ components (see Fig. 3) can be associated with the three sublevels of the U atom, $C_{d_1}$ of the two D atoms, and $C_{d_2}$ and $C_{d_3}$ of the two D atoms, respectively. Thus, we can say that the observed two components of higher binding energy are attributed to the effect of crystal field splitting~\cite{cot} on the Sn 4$d$ core levels of the two D atoms, rather than to different core levels for the two  inequivalent D atoms in the IDA structure~\cite{tej2}. The initial-state theory for the 1U2D (IDA) structure shows that the $C_{d_1}$, $C_{d_2}$, and $C_{d_3}$ sublevels for the two D atoms shift to higher binding energy by 155 (157${\pm}$1), 237 (238${\pm}$2), and 256 (257${\pm}$3) meV, respectively, relative to the average value of three sublevels for the up atom. These shifts are smaller than those (230${\pm}$40 and 390${\pm}$40 meV) of $C_2$ and $C_3$ relative to $C_1$, which were resolved from the high-resolution photoemission spectra~\cite{tej2}. This difference of Sn 4$d$ core-level shifts between the initial-state theory and the photoemission experiment may reflect the final-state screening effects~\cite{peh}.

\begin{figure}[ht]
\centering{ \includegraphics[width=7.0cm]{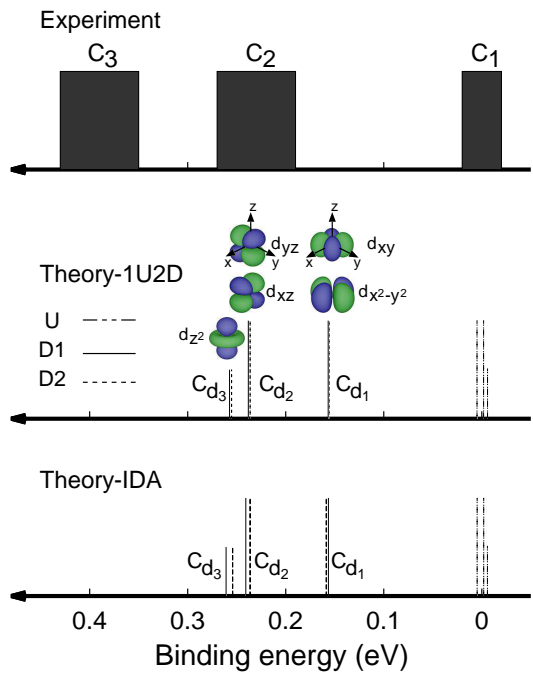} }
\caption{(Color on line) Calculated Sn 4$d$ surface core-level shifts of the 1U2D and IDA structures, in comparison with the high-resolution photoemission experiment~\cite{tej2}. The shifts for the two D atoms (D1 and D2) are given with respect to the average of the three sublevels for the U atom. The positive sign indicates a shift to higher binding energy. The five $d$ orbitals of the D1 atom are displayed with an isosurface of ${\pm}$0.2 (e/{\AA}$^3$)$^{1/2}$.}
\end{figure}

In order to provide an explanation for the observed insulating ${\sqrt{3}}{\times}{\sqrt{3}}$ phase, Profeta and Tosatti~\cite{pro} performed the LDA + $U$ calculation to propose the Mott-Hubbard insulator, where the inclusion of electron correlations strongly modifies the ground state of Sn/Ge(111) from a NM 3${\times}$3 metal to a magnetic insulator with ${\sqrt{3}}{\times}{\sqrt{3}}$ structural symmetry~\cite{pro}. Here, the FI spin order in the Mott-Hubbard insulating state was stabilized within a localized or Heisenberg picture where an unpaired electron was localized at each Sn adatom site with a spin moment of 1 ${\mu}_B$. This localized picture for magnetism contrasts with the present itinerant or Slater magnetism where the magnetic moment is well delocalized over Sn adatoms as well as Ge substrate atoms, as discussed below. Since the Sn DB state is largely delocalized through hybridization with the Ge surface states [see Fig. 2(a)], the SIE inherent to the PBE functional may cause the incorrect prediction of the metallic 3${\times}$3 structure as a ground state. To circumvent such an over-delocalization of Sn DB electrons, we use the HSE+vdW scheme to optimize the NM and magnetic structures of Sn/Ge(111). The calculated total energies of the 1U2D, FI, and ferromagnetic (FM) structures relative to the NM-${\sqrt{3}}{\times}{\sqrt{3}}$ structure are given in Table I. We find that the FI structure is energetically favored over the 1U2D and FM ones by 6.1 and 19.0 meV per ${\sqrt{3}}{\times}{\sqrt{3}}$ unit cell, respectively. The optimized geometry of the FI structure is shown in Fig. 1(b), where the buckling of three Sn atoms within the 3${\times}$3 unit cell is suppressed to become flat, leading to a ${\sqrt{3}}{\times}{\sqrt{3}}$ structural symmetry~\cite{refcom}. In Fig. 2(b), the calculated band structure of the FI structure shows a band-gap opening of 71 meV, in good agreement with the ARPES measurement of 60 meV~\cite{cor}. Therefore, the present HSE+vdW calculation predicts an insulating FI ground state with ${\sqrt{3}}{\times}{\sqrt{3}}$ structural symmetry, consistent with the observed insulating ${\sqrt{3}}{\times}{\sqrt{3}}$ phase~\cite{cor}.

Since the total energy obtained from the HSE+vdW calculation is composed of the HSE energy ($E_{\rm HSE}$) and the vdW energy ($E_{\rm vdW}$), the total energy difference between the 1U2D and FI structures can be divided into the two components, ${\Delta}E_{\rm HSE}$ and ${\Delta}E_{\rm vdW}$. For this decomposition, we obtain ${\Delta}E_{\rm HSE}$ = 4.9 meV~\cite{deltahse}, which is larger than ${\Delta}E_{\rm vdW}$ = 1.2 meV. Thus, we can say that the correction of SIE with the HSE functional gives a more dominant contribution to the stabilization of the insulating FI structure over the metallic 1U2D structure, compared to that from vdW interactions.

Figure 2(b) also shows the charge characters of the spin-up (denoted as $S_{1\uparrow}$ and $S_{3\uparrow}$) and spin-down ($S_{2\downarrow}$) DB states in the FI structure. It is revealed that $S_{1\uparrow}$ and $S_{3\uparrow}$ represent some hybridization of two DB electrons. All of the three DB states strongly hybridize with the $p_z$ orbitals of the surface and subsurface Ge atoms. This strong hybridization gives rise to a large delocalization of spin moments up to deeper Ge atomic layers (see Fig. 4). The sum ($m$) of the spin moments of Sn atoms or Ge atoms in each layer is also given in Fig. 4. Here, the spin moment of each atom is calculated by Mulliken analysis. We find that the Sn layer has $m$ = 0.22 ${\mu}_B$, while the first and third Ge layers have $m$ = 0.40 and 0.22 ${\mu}_B$, respectively, which are significantly larger than those obtained from other Ge layers. Note that the total spin moment is 1 ${\mu}_B$ per 3${\times}$3 unit cell. The result of a large spin delocalization over Sn atoms and Ge substrate atoms contrasts with the case of the previously proposed Mott-Hubbard insulator~\cite{pro} where each Sn atom has a localized spin moment of 1 ${\mu}_B$ as a consequence of electron correlations. It is remarkable that the present FI order is determined by an itinerant single-electron approach with the correction of SIE, thereby representing a Slater insulator driven by itinerant magnetism.

\begin{figure}[ht]
\centering{ \includegraphics[width=7.0cm]{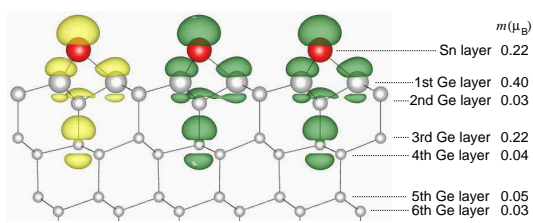} }
\caption{(Color on line) Spin density of the FI structure. The majority (minority) spin density is displayed in dark (bright) color with an isosurface of 0.01 ($-$0.01) e/{\AA}$^3$. The sum ($m$) of the spin moments of Sn atoms or Ge atoms in each layer is also given. For the spin moment of each Ge atom, see Table IS of the Supplemental Material~\cite{supp}.}
\end{figure}

We note that the total energy difference ${\Delta}E$ between the metallic 1U2D structure and the insulating FI structure is 6.1 meV per Sn atom. Although the MIT temperature can be predicted by the precise entropy-related free energy difference ($T$${\Delta}S$) between the 1U2D and the FI structures, we roughly estimate it by considering only the electronic contribution to the entropy, as done by Profeta and Tosatti~\cite{pro}. Assuming that the FI structure has a lack of spin entropy and a charge gap, $T$${\Delta}S$ was approximated to ${\gamma}T^2$ where ${\gamma}$ is the electronic specific heat coefficient (roughly of order ${\sim}$0.1 meV/site K$^2$)~\cite{pro,org}. We thus estimate the MIT temperature as ${\sim}$8 K, which is somewhat below the observed MIT temperature of ${\sim}$30 K~\cite{cor}. This deviation of the MIT temperature may reflect that the 1U2D and FI structures have different vibrational contributions to the entropy, which are not taken into account in $T$${\Delta}S$.

In conclusion, our semilocal and hybrid DFT calculations showed the different predictions for the ground state of the Sn/Ge(111) surface. Contrasting with the PBE functional predicting a metallic 3${\times}$3 ground state, the HSE functional showed that the correction of SIE cures the delocalization error to predict an insulating FI ground state with ${\sqrt{3}}{\times}{\sqrt{3}}$ structural symmetry. We found that the magnetic moment of the FI structure is well distributed over Sn adatoms as well as Ge substrate atoms. It is thus demonstrated that the observed insulating phase in Sn/Ge(111) can be represented as a Slater insulator through itinerant magnetism rather than a Mott-Hubbard insulator driven by Coulomb interactions. We notice that the Sn/Si(111) surface has also been much studied to determine its exact crystallographic arrangement, electronic structure, and ground state~\cite{li,han}. Similar to the present case of Sn/Ge(111), we anticipate that the correction of SIE would be of importance to describe the structural, electronic, and energetic properties of the isoelectronic Sn/Si(111) system.

This work was supported by National Research Foundation of Korea (NRF) grant funded by the Korean Government (NRF-2011-0015754). The calculations were performed by KISTI supercomputing center through the strategic support program (KSC-2012-C3-18) for the supercomputing application research.

\noindent $^{*}$ Corresponding author: chojh@hanyang.ac.kr



\begin{thebibliography}{99}
\bibitem{ctest} Calculations with the k-point sampling of 20${\times}$20 meshes in the surface Brillouin zone of the ${\sqrt{3}}{\times}{\sqrt{3}}$ unit cell, the eight-layer slab of Ge(111), and the force criterion of 0.01 eV/{\AA} change the energy difference between the 1U2D and FI structures by less than 0.2 meV per ${\sqrt{3}}{\times}{\sqrt{3}}$ unit cell. It is also found that the calculated spin moments of Sn and Ge atoms in the FI structure change little by less than 0.01 ${\mu}_B$.
\bibitem{refcom} Since the HSE calculation without vdW interactions predicts the insulating FI ground state with the disappearance of Sn buckling, we can say that the flat structure of Sn atoms is not due to vdW interactions but to the FI spin ordering between Sn atoms. It is found that the Sn-Ge bond lengths and the height difference between the up and down Sn atoms (in the 1U2D structure) obtained using HSE+vdW decrease by less than 0.02 {\AA}, compared to those obtained using HSE.
\bibitem{deltahse} It is notable that ${\Delta}E_{\rm HSE}$ obtained using the HSE calculation without vdW interactions is 2.0 meV. This HSE value is somewhat different from that (4.9 meV) obtained from the HSE+vdW calculation, possibly due to the use of two different Ge lattice constants optimized using HSE and HSE+vdW.
\end{thebibliography}
\end{document}